\documentclass[12pt, prd]{revtex4}
\usepackage{amsmath}
\usepackage{amssymb}

\begin{document}

\title{Why do we live in four dimension?\footnote{First V. V. Narlikar Memorial Lecture delivered on 23 January 2009 at Jamia Millia Islamia, New Delhi.}}
\author{Naresh Dadhich }
\email{nkd@iucaa.ernet.in}
\affiliation{IUCAA, Pune 411007, India}
\begin{abstract}
{\it We perceive the dimension of physical spacetime we live in through physical experiments and hence it is pertinent to probe the dimension in which the fundamental physical forces exist and act? In this context we shall investigate the two classical fields of gravitation and electromagnetism and argue that four dimension is necessary for spacetime but may not be sufficient. Some motivation for higher dimension would also be discussed.}  
\end{abstract}
\maketitle
First of all I wish to warmly congratulate the Vice Chancellor, Jamia Millia Islamia, Professor Mushirul Hasan and his colleaguse in the relevant university bodies on instituting the Memorial Lecture in the honour of Professor V.V. Narlikar in his birth centenary year. Professor Narlikar was one of the two persons, the other being Professor N.R. Sen in Kolkata, who pioneered research in Einstein's general relativity in early 1930s at Benaras Hindu University. He trained a large number of students who spread over other universities and established a school in classical GR. The most notable among them all was Professor P.C. Vaidya, the discoverer  of famous Vaidya solution of a radiating star which is even today very much in vogue  in analysing gravitational collapse and study of singularities. Professor Narlikar was a very erudite lecturer and an inspiring teacher who was held by students and colleaguse alike in awesome esteem and utmost reverence.  
     
I was his doctoral student in his second innings at Pune University as Lokmanya Tilak Professor of Applied Mathematics. There could be nothing more satisfying for a student to give the Memorial Lecture in honour of his teacher. I am very grateful to my friend, Professor M Sami. First, for taking initiative for institution of the Memorial Lecture in the birth centenery year of my revered teacher. It is an apt homage for the man who pioneered research in gravitation which is today one the frontier areas. Second, he has done me a great honour in giving me the opportunity to open the innings of this prestigious series. It is indeed a great privilege and honour and at the same time I feel quite nervous and intimidated by the fear that whether I would be able to live upto my great teacher's daunting standards as well as to the occasion.
   
In this lecture I would follow Professor Narlikar's spirit of story telling who believed that if one understood what she or he was talking about, it should be possible to tell it as a story and the story should expose something new and insightful. I have a formidable benchmark to meet. 
\section{Introduction}
It is a pertinent question to ask why is the physical universe we live in is four dimensional? The universe as we understand seems to be governed by fundamental physical forces. What is the most natural fundamental force and in what dimension does it work? A force, which links to everything that physically exists, in other words an interaction in which all particles participate, would naturally be the most fundamental. It is remarkable that the property of universal linkage would, in a straightforward way, determine the dynamics of the force entirely and would also make a demand on spacetime dimension.
 
In this discourse, we shall focus on basic concepts and principles and shall then attempt to expand and extrapolate on them. We shall begin with a critique of Newton's laws of motion and expound on what could be envisioned as seeded in them? Next it would be demonstrated that the most fundamental universal force is uniquely Einstein's gravity and it requires minimum four dimension for its proper dynamics to be realised. It also turns out that electromagnetic field also lives in four dimension. This shows that four dimension is necessary but is it sufficient? However there are some strong even classical arguments for higher dimension.  
\section{Newton's Laws}
Let us begin with Newton's First Law (NFL) which states that state of no motion (rest) is equivalent to state of uniform motion, constant speed in a straight line. The two are indistinguishable for any physical experiment or observation. This fact by implication means physics remains the same in all frames which are in uniform relative motion, the Principle of Relativity. Further any change from this situation can only occur by application or presence of a force. That is, NFL characterizes the force free state of motion and any departure from it is indicative of its presence.

This is a universal statement which is indepedent of particle parameter and should therefore be true for everything. By universal we shall mean a property or statement which is true for all objects as well as for all reference frames (observers). This should also be true for zero mass particle which obviously cannot be at rest in any frame. It has to move relative to all observers and with a constant speed. Its speed should therefore be a limiting speed for all observers and hence a universal constant. If such a particle exists, it should move relative to all observers with the same constant speed in a straight line. 

Another characterization of absence of force is that space is homogeneous and isotropic and time is homogeneous. Homogeneity of space means motion is completely free of the coordinates and henec they could be freely interchanged, for example, $x \to y$ and vice-versa. Since time is also homogeneous which means motion doesn't, like the space coordinates, depend upon time as well. As we could interchange $x$ and $y$, we should also be able to do the same for $x$ and $t$ for homeogeneity in space as well as in time. That is $x \to t$ but it couldn't be done  because their dimension does not match. If the concept of homogeneity has to be respected, this should happen and we have to match their dimension. That could only be done by asking for a universally constant velocity, $c$, so that $x \to ct$ and vice-versa. Once again we require a constant velocity which could be identified with the velocity of zero mass particle. Space and time are now bound together through this constant velocity and we now have the four dimensional spacetime in place of space and time. NFL characterizes homogeneity of spacetime which asks for a universal constant velocity. 

By homogeneity and universality of NFL, we are naturally driven to the existence of zero mass particles in nature propagating with a universally constant velocity. This is a profound prediction entirely dictated by concept and principle. If we admit them, we need a new mechanics because in Newtonian mechanics velocities add as $w=u+v$ which cannot keep any velocity constant for all observers. The new mechanics couldn't be anything different from Special Relativity (SR). SR could have in principle be discovered independent of and even before Maxwell's electrodynamics. If that were the case, it would have been the most remarkable discovery ~\cite{n1}. That is, first a prediction based on general principles and then its actual verification provided by Maxwell's theory and  universal constant velocity being identified with the velocity of light. In its quantum description, light is the zero mass particle. 

The universal statement that follows from NFL is that motion in absence of 
force (free motion) is always uniform in a straight line. This is a geometric statement and is therefore entirely the property of spacetime geometry without reference to any particle. It is given by geodesic (straight line) of $4$-dimensional homogeneous and isotropic spacetime described by Minkowski metric. This is how it is entirely synthesized in the geometry of spacetime.  

Now we turn to Newton's Second Law (NSL), $m \ddot x^i = F^i$. This equation Professor Narlikar used to interpret as saying, when $F=0$, then either $m \neq 0$ and $\ddot x^i = 0$ or $m=0$ and $\ddot x^i$ indeterminate. The former means there exists a frame relative to which acceleration is zero while the latter indicates absence of a frame and thereby no mass either. This was his way of indicating to Mach's principle which in its simplest avatar surmised, inertia (mass) of a particle arises out of its interaction with other particles in the universe. In the limiting case, there should at least exist one more object and thereby a reference frame to give mass non-zero vaule. Therefore, in absence of frame, particle can't have mass. 

There is yet another possibility. A particle can however exist even when $m=0$ but it has always to move with universal constant velocity. However its motion is now entirely determined by the geometry of spacetime. It is always a geodesic (straight line). That is, motion of zero mass particle is always synthesized in spacetime geometry and no external force can act on it. NSL is therefore not applicable to massless particle at all because even in presence of force, its motion can only be described by geodesic of a suitable spacetime geometry. This is a very important fact which we shall invoke next. 
\section{Universal Force}
We postulate that there exists an interaction in which all particles interact with each other through a universal force. It is universal because of its universal linkage to all particles as well as its presence everywhere and always. It signifies a state dual (opposite) of no force which is never there while it can never be removed. It is always present and cannot be removed from everywhere. As absence of force signifies homogeneity, similarly presence of force implies inhomogeneity. Since force is universal and hence it is always present for all particles, it implies that spacetime has necessarily to be inhomogeneous (and/or anisotropy which would always be implied in this context). Therefore inhomogeneity should be incorporated in the structure of spacetime in the same way as velocity of light is in the geometry of homogeneous spacetime. Geometrically, what distinguishes homogeneous and inhomgeneous space is the Riemann curvature which vanishes for the former and it is termed flat while it is non-zero for the latter and it is then curved. Inhomogeneity thus produces curvature in spacetime. Thus universal force curves spacetime and hence it gets synthesized into spacetime geometry. Its dynamics is now completely determined by the spacetime geometry - its curvature. Alternatively, we could have also argued that its linkage to zero mass particle could be negotiated only if it curved spacetime. Since motion of zero mass particle is not goverened by NSL instead it is always given by geodesic of spacetime, it could only feel universal force through curved spacetime geometry ~\cite{n1,n2}. Then massless as well as massive particles simply follow geodesics of curved spacetime and universal force no longer remains an external force.

We thus have both universal force and no force, which are dual to each-other, are fully incorporated in geometry of spacetime, the former curved while the latter flat. This is very satisfying and what it indicates is the general principle that anything universal should ultimately be synthesized in the spacetime geometry.

Let us now turn to curvature of spacetime and derive from that dynamics of the universal force. It is given by Riemann curvature tensor which satisfies Bianchi differential identity (analogue of curl of gradient, divergence of curl),
the anti symmetric covariant derivative is identically zero, 
\begin{equation}
R^i{}_{m[lk;n]} = 0 . 
\end{equation} 
If dynamics of universal force has to follow from the curvature, it has to follow from this identity which is the only available geometric relation. The only thing we can do to it is to contract on the available indices which does lead, unlike for scalar and vector case, to a non-vacuous relation, 
\begin{equation}
G^{a}{}_{b;a} = 0, ~~~~ G_{ab} = R_{ab} - {1\over2}Rg_{ab} 
\end{equation}
where $R_{ab}$ is the Ricci tensor, the contraction of Riemann, while $R$ is the trace of Ricci. Thus the trace (contraction) of the Bianchi identity yields a non-trivial differential identity from which we can make the following statement  
\begin{equation}
G_{ab} = \kappa T_{ab} - \Lambda g_{ab}, ~~~ T^{a}{}_{b;a} = 0 
\end{equation}  \\
where $T_{ab}$ is the second rank symmetric tensor with vanishing divergence,  the second term is a constant relative to covariant derivative, and $\kappa$ and $\Lambda$ are constants. The left hand side is a second order differential operator on the metric $g_{ab}$ (like $\nabla^2\phi$). For it to become an equation of motion, the tensor $T_{ab}$ should represent the source/charge for force. A source for universal force should also be universal;i.e. something which is shared by all particles and hence it should represent energy momentum distribution and the equation also ensures its conservation. With this identification, the above equation is Einstein's equation for gravitation. We have thus ended with Einstein gravity.

The universal interaction we postulated is in fact nothing but Einstein's gravity and its dynamics entirely follows from the spacetime curvature. This is something very remarkable because no other force makes such a demand on spacetime that it has to fully imbibe its dynamics. For all other forces, spacetime provides a fixed inert background. It is the universality which integrates it with spacetime. Since its dynamics is now property of spacetime, we have no freedom to prescribe a force law, it all follows from the spacetime curvature. Note that Newton's inverse square law is contained in the above equation as a weak field limit.

There are two constants in the equation of which $\kappa$ is to be determined by experimentally measuring the strength of the force and is identified with Newton's constant, $\kappa = -8 \pi G/c^2$. Why is there new constant $\Lambda$ which though arises in the equation as naturally as the energy momentum tensor, $T_{ab}$? It is perhaps because of the absence of fixed spacetime background which exists for the rest of physics and the new constant may be a signature of this fact. It should be noted that homogeneity and isotropy of space and homogeneity of time signifying force free state will in general be described by spacetime of constant curvature and not necessarily of zero curvature. The new constant $\Lambda$ is the measure of the constant curvature of de Sitter (dS) or anti de Sitter (AdS) space. It may in some deep and fundamental sense be related to the basic structure of spacetime. 
\section{Universal Force and Newton's Second Law}  
We have obtained the equation of motion for Einstein's gravity from curvature of spacetime simply by following the differential geometric identity in a straightforward and natural manner. No reference to the celebrated Principle of Equivalence (PE) which served as a great motivation and played cornerstone role in Einstein's journey from SR to General Relativity (GR). It is based on mass proportionality of gravitational force, thereby the universality of acceleration by application of NSL for all massive particles. In the usual equation of motion, we have $m_i \ddot x = m_g \nabla \phi$ where $m_i$ is inertial mass and $m_g$ is gravitational mass. The fact that the acceleration could be anulled out in freely falling lift requires $m_i = m_g$. However, why should that be so and there is no conceivable physical reason for that? Finding the physical reason for the equality is in itself as formidable a problem as one can think of. It is simply being taken as an empirical observational fact without any explanation and understanding. On the other hand, it could be easily bypassed as we didn't have to make any reference to it in deriving Einstein's gravitational equation. It simply followed from spacetime geometry. 

The question is application of NSL to gravity. Since universal force has also to link to massless particle, the latter's equation of motion has to be free of mass. The only way it can happen is when motion is given by geodesic of spacetime geometry which incorporates force in its curvature. Once we have curved spacetime, then its curvature itself dictates the dynamics of the force. Motion under gravity for both massive as well as massless particles is simply described by geodesics of curved spacetime. The universality of acceleration of massive particles is because of geodetic motion in curved spacetime and not because of equality of inertial and gravitational mass.$^{[1]}$\footnotetext[1]{The E$\ddot{o}$tv$\ddot{o}$s like experiments are interpreted to establish their equality to very high precision, one part in $10^{14}$. This could as well be interpreted as the experimental verification of gravity being described by curved spacetime. However in GR, the gravitational field equation (3) which incorporates $m_i = m_g$ because of the universal charcater of gravitational charge, energy-momentum, which is also a measure of inertia.} The question of their equlity therefore does not arise. The point to be noted is that NSL is not applicable for motion under Einstein's gravity. The curved spacetime naturally incorporates PE in the property that at a given point it is always possible to define a tanget plane which is free of gravity giving local inertial frame (LIF). Since spacetime is curved, there can exist no global inertial frame but only LIFs and then Principle of Relativity says all LIFs are equivalent. Thus PE becomes a property of curved spacetime and not so much a driving force for Einstein's gravity. J. L. Synge was the first to voice this sentiment forcefully when he famously pronounced,{\it " The Principle of Equivalence performed the essential office of midwife at the birth of general relativity, but, as Einstein remarked, the infant would have never got beyond its long-clothes had it not been for Minkowski's concept. I suggest that the midwife be now buried with appropriate honours and the facts of absolute space-time faced"} ~\cite{synge}. 

Let us reiterate that motion under gravity like no force is purely a property of spacetime geometry and hence cannot be governed by Newton's laws of motion. This is an important feature not often emphasized upon. 
\section{Dimension of Spacetime}
Dimension of spacetime we perceive only through physical experiments and hence it is pertinent to ask the question in what dimension physical fields live? Let  us begin with gravity. Its fundametal entity is Riemann curvature which requires minimum 2 dimension for its definition. However it is obvious that for any physical phenomenon, there should at least be one space and one time dimension. Two dimension is therefore required without reference to any interaction. Since gravity is present everywhere and hence it should have a massless free propagation;i.e. Einstein equation should admit non-trivial vacuum solution allowing for free propagation. It turns out that in 2 and 3 dimension, number of Ricci and Riemann components are the same and hence it cannot admit non-trivial vacuum solution for free propagation. That is, 2 and 3 dimension are not big enough to incorporate free propagation. So we come to 4 dimension where Riemann has 20 components while Ricci has 10. Hence it admits non-trivial vacuum solution and thereby free propagation through gravitational wave.  

The other classical field is Maxwell's electromagnetic field which unlike gravity links to a specific bipolar electric charge, however retaining the long range property. It is described by a gauge vector field. Since it is a long range force, it has also to have, like gravity, massless free propagation. It is easy to see from Maxwell's equations that this also cannot happen in dimension $<4$. 

Thus both gravity and electromagnetic fields require minimum 4 dimension for their dynamics. This means 4 dimension is necessary but is it sufficient too? Why not dimension $>4$? To probe this question further, we have to ask the question, is there any property, like the free propagation, of either of the fields which cannot be accommodated in 4 dimension? We have to identify that. 

For electromagnetic field, there seems to be no property which remains unaddressed. It however obeys an interesting property of the scale or conformal invariance. That is, the action Lagrangian $F_{ik}F^{ik}\sqrt{-g}$ remains invariant under the confromal transformation  $g_{ab} \to f^2g_{ab}$ only in 4 dimension. If we make conformal invariance as an abiding principle, electromagnetic field can only live in 4 dimension. Clearly gravity cannot be conformally invariant simply because the metric for it is a dynamical variable. In contrast for the rest of physics it simply defines the spacetime background but does not participate in dynamics of the interaction. For other than gravity, it appears natural that physics should not change when scale is unversally changed. However, this will not be true for a field which has an inherent scale like the weak field or any massive field. A scale is introduced by spontaneous symmetry breaking and the resulting theory may be an intermediate effective theory. In the complete theory, the symmetry may be restored and so would conformal invariance. May what the ultimate situation be, it is undeniable that conformal invariance is aesthetically very appealing and satisfying.   

Back to the main question of yet unexplored property. For electrodynamics, it has not been possible to tag any such porperty.$^{[2]}$\footnotetext[2]{The other two basic forces are purely quantum fields which we understand very little at the intuitive level and hence it is very difficult to subject them to this kind of general consideration. We would therefore not indulge into their discussion.} We could probe this question somewhat meaningfully for gravity and that is what we shall now do. We shall however attempt to identify three features of gravitational field whcih seem to have remeined unattended so far. These are purely classical motivations for higher dimension ~\cite{n1,n3,n4}.

(a) Flat Space Embedding: Does gravity remain confined to 4 dimension is equivalent to asking does curvature remain confined to a given dimemsion? At first sight, the question appears uncalled for because curvature is intrinsic to a spacetime of given dimension and hence where is the question of its not remaining confined to it? That is true but in the question at hand, curvature embodies dynamics of field, it is then not out of place to ask, does gravitational dynamics propagate in higher dimension or not? What could be a test for whether it does or does not? One such test perhaps could be isometric embedding in $5$-dimensional flat spacetime. If that happens, gravity has not propagated out.  There is a theorem in differential geometry which addresses precisely this question and it states that $n$-dimensional curved space requires minimum $n(n+1)/2$ dimension for its flat space embedding. This means $4$-dimesnional gravity can in general leak down to 10 dimension! The prototype gravitational field of a masspoint described by Schwarzschild solution requires 6 dimension for its flat space embedding while conformally flat FRW metric is indeed embeddable in $5$-dimensional flat spacetime. Since FRW has vanishing Weyl curvature which means there is no free gravity to propagate out any further! 

(b) Self Intercation: Gravity is an inherently self interactive force and the self interaction could only be evaluated by successive iterations. Einstein's gravity includes self interaction but only in first iteration through square of first derivative of metric in Riemann curvature. The question is how do we stop at the first iteration or do we have any physical justification for not going any further? There seems to be no such physical reason and hence it is imperative to go to next iterations. The second iteration would ask for a quadratic polynomial in Riemann curvature which should give the corresponding term in the equation of motion. This will square the second derivative as well. We would however like to have the second order quasilinear (highest order of derivative to be linear) equation which will only happen with the specific coefficients in the polynomial, known as Gauss-Bonnet term. Alternatively, we could ask for a fourth rank tensor which is a homogeneous quadratic polynomial in Riemann curvarture and whose anti symmetric Bianchi derivative on contraction yields a divergence free second rank symmetric tensor (analogue of Einstein  tensor). This will also identify Gauss-Bonnet and in general Lovelock polynomial ~\cite{bia}. The remarkable property of Gauss-Bonnet polynomial is that it makes no contribution in the equation of motion for dimension $<5$. Thus we have to go to higher dimension for physical realization of second iteration of self interaction. Remaining stay put in 4 dimension, we are neglecting the second iteration of self interaction which might become significant at higher energies. Therefore it should certainly be relevant for quantum gravity realm.

We can thus say like 2 and 3 dimension were not big enough for free propagation similarly 4 dimension is not big enough to fully accommodate self interaction dynamics of gravity. The question is, how far do we go this way. If we envision that matter remains confined to $3$-brane (i.e. $4$-spacetime) which is also the case for string theory, bulk spacetime is therefore completely free of matter and hence it could only have constant curvature. The $3$-brane bounds the bulk which is free of matter and hence of constant curvature and it could be dS/AdS. It has zero Weyl curvature and so there is no free gravity to propagate any further. We seem to end up with a scenario with dS/AdS bulk being bounded by $3$-brane harbouring matter, quite similar to the AdS/CFT picture ~\cite{mal}. It is also similar to Randal-Sundaram brane world gravity matter on $3$-brane and AdS bulk ~\cite{rs}. In this construction the iteration naturally stops at second level. This is however not a general setting and hence the question of how far to go in iteration remains pertinent and open. 

(c) Charge Neutrality: For a classical field, it is natural to ask for total charge to be zero. How could this happen for gravity because its source matter/energy is always positive - unipolar? The only way it could be balanced is by field having charge of opposite polarity - field energy being negative. It is however distributed all over the space and hence not localizable. This leads to the simple and natural explanation for why gravity or rather universal force is always attractive because total charge has to be zero. This is what was rigorously demonstrated by the famous ADM calculation that if one integrates gravitational field energy surrounding a mass point for the whole of space, it would exactly balance the mass ~\cite{adm}. If an infinitely dispersed distribution of bare mass $m$ is let to collapse under its own gravity, the ultimate end result would be field entirely eating up mass $m$ as the centre $r=0$ is approached. The vanishing of gravitational Hamiltonian also indicates in certain sense the fact that 'total charge' is zero for gravity.  

Let us now apply this property to field of a masspoint at rest somewhere. In its finite neighbourhood, there would be over dominance of positive charge because some negative charge of field energy has been left out. Whenever charge is not fully balanced on a surface, field must propagate off it;i.e. gravity must propagate in extra dimension. However the strength of its charge goes on diminishing because as it propagates its past light cone goes on encompassing more and more of negative charge (field energy). It propogates in extra dimension but with diminishing charge strength and hence not deep enough. This is quite analogus to the rough intuitive picture one has in the case of strong force where the coupling becomes stronger with distance keeping quarks confined and there is asymptotic freedoem at $r=0$ end. The picture that emerges is that zero mode free propagation remains completely confined to $3$-brane while propogation in extra dimensional bulk has sharp fall off and hence it also remains confined to the brane neighbourhood. The scenario that emerges is quite similar to Randal-Sundaram brane world gravity where zero mode remains confined to the brane and bulk is AdS, spacetime of constant curvature ~\cite{rs}. 

Gravity may therefore propagate in higher dimensional bulk but not as a free field but with diminishing charge strength and hence not deep enough. 
\section{Beyond Four Dimension}
We have seen above there are reasonably persuasive physical arguments inspired by classical features of Einstein's gravity for higher diemsnion. Higher diemsnion is however natural arena for string theory and quantum gravity approach emanating from it. It may be worth noting that one loop correction in string theory does generate Gauss-Bonnet term. Perhaps it indicates that subsquent order loop corrections may give rise to next order terms in Lovelock polynomial. If that happens, it would be very exciting. The order of iteration of self interaction seems to correspond with order of loop correction. The former is tied to dimension which should be $>4$ for the second iteration. It is natural to envisage that higher order iteration of self interaction is realised at higher energy and in higher dimension. Further its correspondence to order of loop correction which is undoubtedly a  higher energy effect is not a coincidence but instead it may be strongly indicative  of the fact that Gauss-Bonnet term and in general Lovelock polynomial represent an intermidiary state between classical and quantum gravity. Further its physical realization naturally asks for higher dimension. The order of iteration for self interaction is therefore like loop correction remains open. It is the scale of energy which would determine how deep it is able to fathom dimension.   

One of the obvious questions is if there exists
 higher dimension, why don't we see and experience it? How do we 'see' dimension? We do that through physical experiments and for that most common reliable probe is electromagnetic interaction which remains confined to 4 dimension. It cannot probe dimension $>4$. We have to resort to a field which can propagate in higher dimension. That's only gravity and that too in a manner (non free propagation) which we are not familiar with. To probe higher dimension, we have therefore to devise a purely gravitational experiment and also a method to fathom its unfamiliar propagation in higher dimension. This is a formidable task. It is hoped that higher dimension would perhaps leave some imprint at sub millimetre scale. The tools are being sharpened to break the millimetre barrier. On the other hand, cosmology is a favourite playground for testing new and exciting ideas and concepts. The cosmological signature of higher dimension is vigorously being pursued through various brane world as well as other very interesting and exotic scenarios. The search is on and it is quite exciting. 

Finally, if I have been able to convey the excitement of understanding things in one's own way leading to some new perspective and insight, I would consider that I have succeeded in paying a worthy and respectful homage to my teacher and of which hopefully neither of us has to feel shy of.    



\end{document}